\documentclass[prl, twocolumn, superscriptaddress]{revtex4-1}
\usepackage{bm, amsmath, amsfonts, amssymb}
\usepackage[italicdiff]{physics}
\usepackage{braket}
\usepackage{comment}
\usepackage{subfigure}
\usepackage{color}
\usepackage{graphicx}

\usepackage{qcircuit}
\usepackage{url}
\usepackage{bm}
\usepackage[breaklinks=true]{hyperref}

\begin{document}
\title{Time-crystalline long-range order in chiral fermionic vacuum}

\author{Nobuyuki Okuma}
\email{okuma@hosi.phys.s.u-tokyo.ac.jp}
\affiliation{
 Yukawa Institute for Theoretical Physics, Kyoto University, Kyoto 606-8502, Japan
}

\date{\today}
\begin{abstract}
It is widely believed that there is no macroscopic time-crystalline order in the ground states of short-range interacting systems.
In this paper, we consider a time-dependent correlation function for an order operator with a spatially discontinuous weight in a one-dimensional chiral fermionic system.
Although both the Hamiltonian and the order parameter are composed of spatially local operators, the time-dependent correlation function diverges logarithmically in equal time intervals.
This result implies a breakdown of an inequality that claims the absence of time-crystalline long-range order in the ground states, unless the upper-bound constant is set to be infinity.
This behavior is due to the combination of the discontinuity of the order operator and the infinite dimensionality of quantum field theory.
In the language of bosonization, 
it can also be related to the divergence of a space-time-resolved bosonic correlation function.

\end{abstract}

\maketitle
The concept of a time crystal was originally proposed by Wilczek as a system that spontaneously breaks the time translation symmetry \cite{Wilczek}.
However, several papers discussed the difficulties of time crystalline orders in ground states \cite{Bruno-13,nozieres2013time,Watanabe-Oshikawa}.
For example, if we restrict our discussion to a Heisenberg operator $\Phi(t)=e^{iHt}\Phi e^{-iHt}$, where $H$ is a time-independent Hamiltonian and $\Phi$ is an observable operator, time-dependent oscillations are trivially forbidden because the expectation value of any Heisenberg operator for a ground state $|\mathrm{GS}\rangle$ does not depend on time \cite{Watanabe-Oshikawa}:
$\langle\mathrm{GS}|\Phi(t)|\mathrm{GS}\rangle=e^{iE_0t}\langle\mathrm{GS}|\Phi|\mathrm{GS}\rangle e^{-iE_0t}=\langle\mathrm{GS}|\Phi|\mathrm{GS}\rangle$, where $E_0$ is the ground-state energy.

Long-range order has also played a central role in the physics of spontaneous symmetry breaking \cite{Tasaki}.
While spontaneous symmetry breaking is only defined for symmetry-broken ground states, long-range order can describe an ordered structure in symmetric ground states.
In this context, the time-dependent long-range order described by a correlation function $\langle \mathrm{GS}|\Phi(t)\Phi|\mathrm{GS}\rangle$ has also been discussed \cite{Watanabe-Oshikawa}.
For macroscopic systems, the order parameter $\Phi$ is chosen as the summation or the integration of local operators over the spatial coordinate, such as the total spin operator in spin systems.
Unfortunately, attempts to find time-dependent long-range orders in the ground states of short-range interacting systems are not so promising.
Watanabe and Oshikawa proved the following inequality for the time-dependent correlation function \cite{Watanabe-Oshikawa}:
\begin{align}
    \frac{1}{L^{2d}}\left|\langle \mathrm{GS}|\Phi(t)\Phi|\mathrm{GS}\rangle-\langle \mathrm{GS}|\Phi\Phi|\mathrm{GS}\rangle \right|\leq A\frac{t}{L^{d}},\label{inequality}
\end{align}
where $L$ is the system size, $L^d$ is the $d$-dimensional volume, and $A>0$ is a constant that only depends on the Hamiltonian and the order operator.
This inequality indicates that the time-dependence of the volume-normalized correlation function goes to zero in infinite-volume limit \cite{Watanabe-Oshikawa}.
Owing to this inequality, the ground-state time crystal is believed to be impossible. Other possibilities for time crystals have been extensively studied in nonequilibrium states of matter
\cite{Sacha-15,Khemani,Else,Yao,zhang2017observation,choi2017observation,Sacha_2017,sacha2020time,else2019discrete,khemani2019brief} and the ground states of long-range interacting systems \cite{Kozin,okuma2021timecrystalline}, while Ref. \cite{guo2021quantum} defined generalized time crystals in ground states.

Strictly speaking, there remains a possibility for a time-crystalline order in the ground states of short-range systems because the finiteness of the constant $A$ in the inequality (\ref{inequality}) was implicitly assumed in previous works. 
In this work, we consider a time-dependent correlation function for an order operator with a spatially discontinuous weight in a one-dimensional chiral fermionic system.
Although the Hamiltonian and the order operator are composed of spatially local operators, we show that the time-dependent correlation function diverges logarithmically in equal time intervals and breaks the inequality ($\ref{inequality}$) for any finite constant $0<A<\infty$.

\begin{figure}
\begin{center}
 \includegraphics[width=8cm,angle=0,clip]{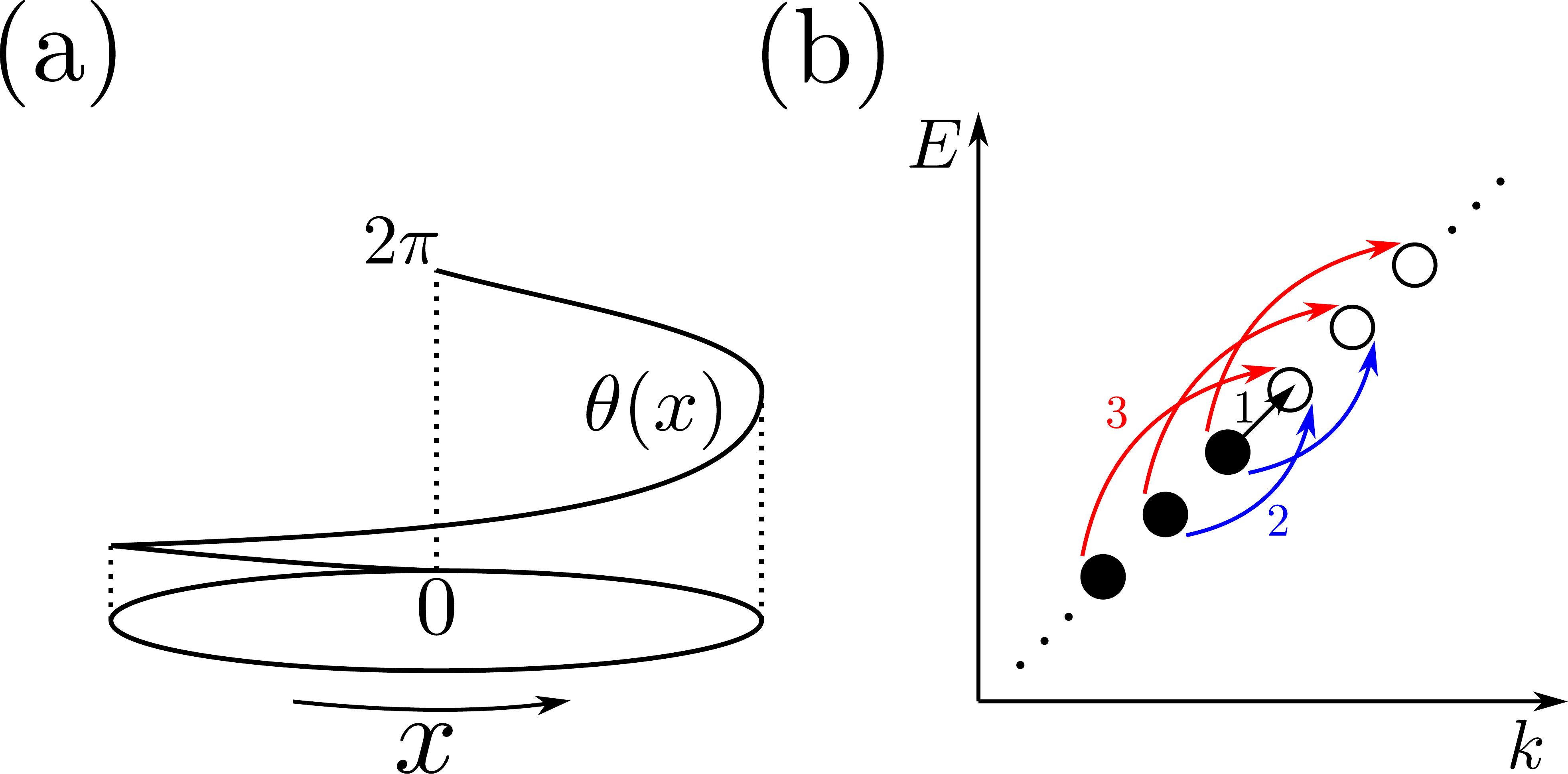}
 \caption{Schematic pictures of (a) weight function of order operator and (b) one-particle excitations from the ground state in chiral fermionic system. }
 \label{fig1}
\end{center}
\end{figure}

\paragraph{Model.---}
We consider a one-dimensional Hamiltonian:
\begin{align}
    H=\int_{0}^L dx \psi^{\dagger}_x(-i\partial_x)\psi_x,
\end{align}
where $L$ is the system size, $x\in[0,L]$ is the spatial coordinate, and $(\psi^{\dagger},\psi)$ are the fermionic creation/annihilation operators. 
The group velocity is set to be unity.
This model describes a free chiral fermionic system without long-range interactions.
By imposing the periodic boundary condition $(x=0\equiv L)$, one can rewrite the Hamiltonian as
\begin{align}
    H&=\sum_{k}k\psi^{\dagger}_k\psi_k,\\
    \psi_k&=\frac{1}{\sqrt{L}}\int_{0}^{L} dx e^{-ikx}\psi_x,
\end{align}
where $k=2\pi m/L$ with $m\in\mathbb{Z}$.
The ground state is simply given by $|\mathrm{GS}\rangle=\prod_{k\leq0}\psi^{\dagger}_{k}|0\rangle$ with $|0\rangle$ being the Fock vacuum.
We set the ground-state energy to be zero.
\paragraph{Divergence of time-dependent correlation function.---}
As an order parameter, we consider the following operator:
\begin{align}
    \tilde{\Phi}:=&\int_{0}^{L}dx\theta(x)(\psi^{\dagger}_x\psi_x-\langle\mathrm{GS}|\psi^{\dagger}_x\psi_x|\mathrm{GS}\rangle )\notag\\
    =:&\Phi-\langle\mathrm{GS}|\Phi|\mathrm{GS}\rangle,\label{orderpara}
\end{align}
where $\theta(x):=2\pi x/L$ is the weight function, and tilde denotes the subtraction of the vacuum contribution.
While this order operator consists of spatially local operators, the weight function is discontinuous at $x=0\equiv L$ under the periodic boundary condition [Fig. \ref{fig1} (a)].
The time-dependent correlation function is expressed as 
\begin{align}
    C(t):=\langle\mathrm{GS}|\tilde{\Phi} e^{-iHt}\tilde{\Phi}|\mathrm{GS}\rangle
    =\left[\sum_{m>0}e^{-iE_m t}|\langle m|\Phi|\mathrm{GS}\rangle |^2\right],
\end{align}
where $E_m$ denotes the excited energy-eigenstates.
We have used $\langle \mathrm{GS}|\tilde{\Phi}|\mathrm{GS}\rangle=0$.
The single-particle spectrum is quantized, and the excited energies are multiples of the first excited energy $E_1=2\pi/L$, which ensures the periodicity of $C(t)$.
The matrix elements take nonzero values for the excited states that are created by exciting one particle from the vacuum.
The relevant excited states with the energy $E_m=l\times E_1$ have the $l$-fold degeneracy [Fig. \ref{fig1} (b)].

\begin{figure}
\begin{center}
 \includegraphics[width=8cm,angle=0,clip]{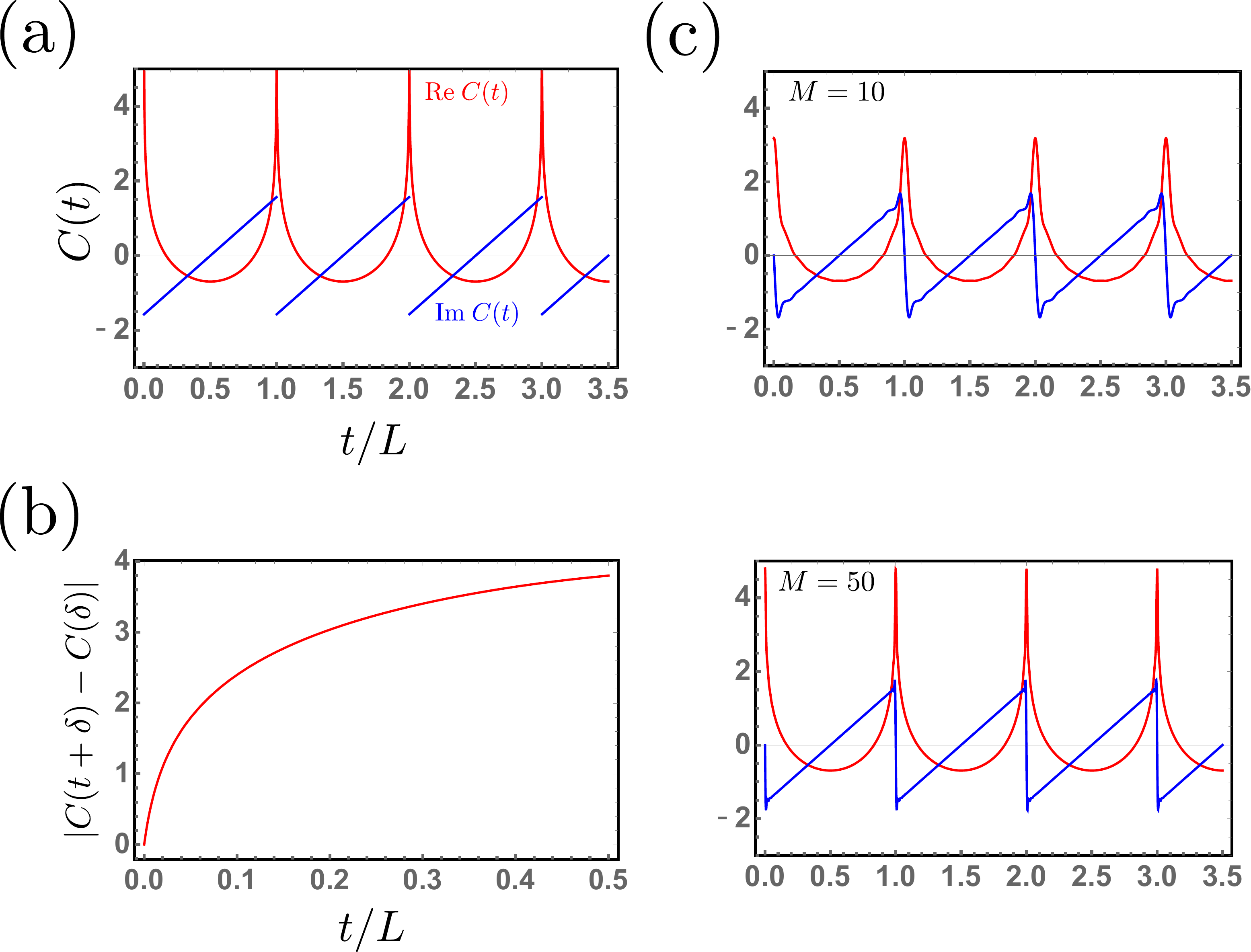}
 \caption{Time-dependent correlation functions. (a) Real and imaginary parts of $C(t)$. Real part diverges logarithmically at equal intervals. (b) The difference between time-dependent correlation functions at different time. The time-scale cutoff $\delta$ is $0.01L$. (c) Time-dependent correlation functions under momentum cutoff $M=10$ and $50$.  }
 \label{fig2}
\end{center}
\end{figure}

The matrix element for an excited state $|l,k\rangle:=\psi^{\dagger}_{k+2\pi l/L}\psi_k|\mathrm{GS}\rangle$ is calculated as 
\begin{align}
    &\langle l,k|\Phi|\mathrm{GS}\rangle=\int_{0}^{L}dx\theta(x)\langle l,k|\psi^{\dagger}_x\psi_x|\mathrm{GS}\rangle\notag\\
    &=\frac{1}{L}\int_{0}^{L}dx \frac{2\pi x}{L}e^{-i\frac{2\pi l}{L}x}\notag\\
    &=\frac{2\pi}{L^2}\left\{\left[  \frac{Lxe^{-i\frac{2\pi l}{L}x}}{-i2\pi l}\right]^{L}_0-\int^L_0 dx \frac{Le^{-i\frac{2\pi l}{L}x}}{-i2\pi l}\right\}=\frac{i}{l}.\label{matelecal}
\end{align}
By taking into account for the $l$-fold degeneracy, we obtain
\begin{align}
    &C(t)=\sum^{\infty}_{l=1}e^{-i\frac{2\pi l}{L}t}l*|i/l|^2=\sum^{\infty}_{l=1}\frac{1}{l}e^{-i\frac{2\pi l}{L}t}\notag\\
    &=-\log \left[1-e^{-i\frac{2\pi }{L}t}\right]=-\log\left[2\sin \frac{\pi t}{L}\right]+i \left[\frac{-\pi}{2}+\frac{\pi \hat{t}}{L}\right],\label{infinitesum}
\end{align}
where $0\leq\hat{t}\leq L$ is equivalent to $t$ modulo $L$.
The period of oscillation $L$ coincides with the time for a chiral particle to circle the system.
Remarkably, the time-dependent correlation function diverges logarithmically at $t=n L/2 \pi$ [Fig. \ref{fig2}(a)].
Thus, the following holds:
\begin{align}
C(t)-C(0)=
\begin{cases}
0&(t=nL,n=0,1,\cdots)\\
\infty & (\mathrm{otherwise})
\end{cases}.\label{diverge}
\end{align}
This implies the breakdown of the inequality (\ref{inequality}) for any finite constant $0<A<\infty$.
Since $\lim_{L\rightarrow\infty}C(t)/L^2=0$ for $t\neq nL$, the following holds in the infinite-volume limit:
\begin{align}
\lim_{L\rightarrow\infty}\frac{C(t)-C(0)}{L^2}=
\begin{cases}
0&(t=nL,n=0,1,\cdots)\\
\infty&(\mathrm{otherwise})
\end{cases}.
\end{align}
Contrary to the expectation in Ref. \cite{Watanabe-Oshikawa}, the volume-normalized time-dependent correlation function depends on time in infinite-volume limit. 
Note that the result does not depend on the chirality of fermions. If there exist both the left- and right-movers, the result becomes simply doubled.

\paragraph{Interpretation by bosonization.---}
One can also derive the same result by using bosonization.
Let us consider the following bosonic order operator:
\begin{align}
    \Phi_B:=&\int_{0}^{L}dx\theta(x)\partial_x\phi_x.\label{bosonorder}
\end{align}
The bosonic field operator $\phi$ is related to the fermionic field operator via a conventional bosonization procedure \cite{von1998bosonization}:
\begin{align}
    \phi_x&=-\sum_{l>0}\frac{1}{2\pi\sqrt{l}}(e^{iqx}b_q+e^{-iqx}b_q^{\dagger}),\\
    b^{\dagger}_q&=\frac{i}{\sqrt{l}}\sum_{k}\psi^{\dagger}_{k+q}\psi_k,
\end{align}
where $q=2\pi l/L$, $(b^{\dagger}_q,b_q)$ are creation/annihilation operators of bosons with energy $q$. 
This order operator is equivalent to the fermionic order operator (\ref{orderpara}) up to a phase factor.
For an excited state $|q\rangle:= b^\dagger_q|0\rangle$ with $|0\rangle$ being the bosonic vacuum, the matrix element is calculated as
\begin{align}
    \langle q|\Phi_B|0\rangle&=\int_0^Ldx\theta(x)\frac{1}{2\pi\sqrt{l}}i\frac{2\pi l}{L}e^{-i\frac{2\pi l}{L}x}
    \langle q|b^{\dagger}_q|0\rangle=\frac{-1}{\sqrt{l}}.
\end{align}
Since the relevant excited states have no degeneracy in the bosonization, one obtains the same expression as Eq. (\ref{infinitesum}).

The more striking fact is that the time-dependent correlation function can be related to the space-time-resolved correlation functions for bosonic field operators, as shown below.
The order operator (\ref{bosonorder}) can be rewritten as
\begin{align}
    \Phi_B&=\left[\theta(x)\phi_x  \right]^{L}_0-\int^L_0 dx \partial_x\theta(x)\phi_x\notag\\
    &=2\pi\left(\phi_{x=0}-\frac{1}{L}\int^L_0 dx\phi_x\right).
\end{align}
By using this expression, one can rewrite the time-dependent correlation function in terms of the bosonic correlation functions:
\begin{align}
    &C_B(t)/(4\pi^2)=\langle0|\Phi_B e^{-iHt}\Phi_B|0\rangle/(4\pi^2)\notag\\
    &=\langle0|\phi_{x=0}(t)\phi_{x=0}|0\rangle+\int_0^L\frac{dx}{L}\int_0^L\frac{dx'}{L}\langle0|\phi_{x}(t)\phi_{x'}|0\rangle\notag\\
    &-\int_0^L\frac{dx}{L}\langle0|\phi_{x=0}(t)\phi_{x}|0\rangle-\int_0^L\frac{dx}{L}\langle0|\phi_{x}(t)\phi_{x=0}|0\rangle.
\end{align}
According to Ref. \cite{von1998bosonization}, the bosonic correlation function for $t\geq 0$ is given by
\begin{align}
    -g(t,x):=\langle0|\phi_{x}(t)\phi_{x=0}|0\rangle=-\frac{1}{4\pi^2}\log (1-e^{-i\frac{2\pi (t+x)}{L}}).
\end{align}
By using the translation invariance of the Hamiltonian and $\int_0^Ldx g(t,x)=0$,
one obtains
\begin{align}
    C_{B}(t)=-4\pi^2g(t,x=0)=C(t).
\end{align}
Thus, the logarithmic divergence of the time-dependent correlation function is reduced to that of $g(t,x=0)$.
Physically, this space-time-resolved correlation function describes the periodically-rotating dynamics of one boson initially placed at the specific point $x=0$.
In other words, the many-body phenomenon in the fermionic language is related to the one-particle dynamics via the bosonization. 

Thus far, we have discussed the divergence of the time-dependent correlation function in a chiral fermionic vacuum.
In the following, we extract finite values from the diverging function by setting the cutoff to time or momentum. 

\paragraph{Inequality under time-scale cutoff.---}
We here consider $C(t+\delta)-C(\delta)$ with $\delta>0$ instead of Eq. (\ref{diverge}).
This shift enables us to extract a finite value from the diverging quantity.
We plot this function for $\delta/L=0.01$ in Fig. \ref{fig2} (b).
In conventional cases, the proof of the inequality (\ref{inequality}) in Ref. \cite{Watanabe-Oshikawa} can be trivially generalized to the case with time shift:
\begin{align}
    \frac{1}{L^{2d}}\left|\langle \mathrm{GS}|\Phi(t+\delta)\Phi|\mathrm{GS}\rangle-\langle \mathrm{GS}|\Phi(\delta)\Phi|\mathrm{GS}\rangle \right|\leq A\frac{t}{L^{d}}.
\end{align}
In the present case, however, there may exist $\delta$ and $t$ for any finite constant $0<A<\infty$ such that
\begin{align}
    |C(t+\delta)-C(\delta)|>ALt.
\end{align}
Instead of a rigorous proof, we consider an approximate analysis under the limit $t,\delta\ll L$.
Under this limit, one obtains
\begin{align}
    |C(t+\delta)-C(\delta)|\simeq \sqrt{\left\{\left[\log \frac{t+\delta}{\delta}\right]^2+\left[\frac{\pi t}{L}\right]^2\right\}}.
\end{align}
Thus, the comparison between the left- and right-hand sides of the inequality is equivalent to that between $t$ and $f(t):=\delta\times (e^{\sqrt{A^2L^2-\pi^2/L^2}t}-1)$.
Since $f(0)=0$, $f'(t)\geq 0$ for $t>0$, and $f'(0)\leq1$ for $\delta\leq 1/\sqrt{A^2L^2-\pi^2/L^2}$, 
the inequality is broken for $\delta<1/\sqrt{A^2L^2-\pi^2/L^2}$ for a while.
In other words, for a function $g(L)$ that decays faster than $1/L$ for $L\rightarrow \infty$, there exists $L_0$ such that the inequality is broken for $\delta<g(L\geq L_0)$ for a while.

\paragraph{Inequality under momentum cutoff.---}
Another way for truncation is to introduce momentum cutoff.
We here introduce the momentum cutoff $\Lambda=2\pi M/L$ to the Hamiltonian ($-\Lambda\leq k\leq\Lambda$).
Under this cutoff, the time-dependent correlation function is given by
\begin{align}
    C_{M}(t)=\sum^{M}_{l=1}\frac{1}{l}e^{-i\frac{2\pi l}{L}t}+\sum^{2M}_{l=M+1}\frac{2M-l}{l^2}e^{-i\frac{2\pi l}{L}t}.
\end{align}
We plot this function for $M=10$ and $50$ in Fig. \ref{fig2} (c).
For large $M$, the dominant term of $C_M(0)$ is a logarithmic function of $M$:
\begin{align}
    C_M(0)&=2H_M-H_{2M}+\sum_{l=M+1}^{2M}\frac{2M}{l^2}\notag\\
    &\simeq\log M+\gamma-\log2,\label{momcutoff}
\end{align}
where $H_M=\sum^M_{l=1}1/l$ is the harmonic number, and $\gamma$ is the Euler’s constant.
We have used an approximate relation $H_M\simeq \log M+\gamma$ and the convergence of $\sum_{l}1/l^2$.
Similarly, under the small time limit $Mt/L\rightarrow 0$, we obtain
\begin{align}
    C_M(t)-C_M(0)\rightarrow&\frac{-i2\pi t}{L}\left[\sum^{M}_{l=1}-\sum^{2M}_{l=M+1}+2M\sum^{2M}_{l=M+1}\frac{1}{l}\right]\notag\\
    \simeq&(-i4\pi\log2 )\frac{Mt}{L}.\label{cutoff}
\end{align}
Under the momentum cutoff, $M$ should diverge faster than $L^2$ in order to break the inequality (\ref{inequality}).
Typically, the momentum cutoff $M$ for the chiral fermionic system in solid-state physics cannot be over $\mathcal{O}(L)$, which indicates that the breakdown of the inequality does not occur in solids.

\paragraph{Generalization to other weight functions.---}
For the divergence of the time-dependent correlation function, the infinite dimensionality of quantum field theory and the discontinuity of the order operator play important roles.
The infinite sum in Eq. (\ref{infinitesum}) is a consequence of the infinite dimensionality of the fermionic vacuum.
In addition, the divergence of the infinite series comes from the slow decay of the matrix element under $l\rightarrow\infty$. 
Actually, one can find the divergence of the time-dependent correlation function for other order operators.
Let us consider 
\begin{align}
    \tilde{\Phi}^\Theta:=&\int_{0}^{L}dx\Theta(x)(\psi^{\dagger}_x\psi_x-\langle\mathrm{GS}|\psi^{\dagger}_x\psi_x|\mathrm{GS}\rangle ),\label{general}
\end{align}
where $\Theta(x)$ is a weight-function defined on $x\in[0,L)$. Equation (\ref{matelecal}) implies that the matrix element is nothing but the Fourier coefficient $\Theta_l$, and the time-dependent correlation function is calculated by the inverse Fourier transform of a coefficient that depends on $\Theta_l$:
\begin{align}
    \Theta_l:=&\frac{1}{L}\int_{0}^{L}dx \Theta(x)e^{-i\frac{2\pi l}{L}x},\\
    C(t)=&\sum^{\infty}_{l=1}l|\Theta_l|^2e^{-i\frac{2\pi l}{L}t}.
\end{align}
For the divergence at $t=nL$, $\Theta_l$ should decays slower than $1/l^{3/2}$ in $l\rightarrow\infty$.
Correspondingly, the function $\Theta(x)$ should contain large discontinuity such as $\theta(L)-\theta(0)=2\pi$ in Eq. (\ref{orderpara}).

A discontinuous weight is also useful to describe subsystem dynamics.
Let us consider the following weight function:
\begin{align}
\Theta(x)=
\begin{cases}
1 & (0\leq x\leq L_{\rm sub}\ll L)\\
0 & (\mathrm{otherwise})
\end{cases}.
\end{align}
When we regard the region $0\leq x\leq L_{\rm sub}$ as a subsystem, the order operator is the total number operator of the subsystem.
By repeating a similar calculation, we obtain $C(t)=-2g(t,0)+g(t,L_{\rm sub})+g(t,-L_{\rm sub})$. 

\paragraph{Discussion.---}
In this paper, we have shown that the inequality (\ref{inequality}) for a finite constant $A$ can be broken by the combination of discontinuity and infinite dimensionality.
In condensed matter physics, however, the breakdown would not occur because there seems to be no doubt about the proof in the case of lattice systems \cite{Watanabe-Oshikawa,watanabe2020proof}.
In the present example, the difficulty for breaking the inequality comes from the fact that the momentum cutoff $M$ cannot be over the system size $L$, as discussed above.
This is because the number of the relevant states in interest is proportional to the volume at most, which is nothing but an extensive nature.
In the case of quantum field theory, however, infinite degrees of freedom can exist per unit volume, which enables us to observe the divergence.
The infinite dimensionality can also appear in bosonic systems such as quantum optical systems, and there is a possibility for the macroscopic time-dependent long-range order.
In our previous work \cite{okuma2021timecrystalline}, we have discussed a macroscopic oscillation in a squeezed ground state. The amplitude of the oscillation diverges in the infinite squeezing limit.
This is not a breakdown of the inequality because the divergence is explained by the fact that $A$ explicitly depends on the Hamiltonian with a diverging parameter.




Apart from the essential breakdown of the inequality in quantum field theory, we note that the inequality can be superficially broken even under a condensed-matter cutoff when the local operators in the order operator and/or the Hamiltonian have implicit or explicit size-dependence.
More precisely, the assumption that the constant $A$ does not depend on $L$ is broken in such cases.
For example, let us consider $\theta(x)=x$ instead of $\theta(x)=2\pi x/L$.
Physically, the new order operator represents the total charge polarization of the ground state.
Thanks to the factor $L$, the time-dependent correlation function with momentum cutoff seems to break the inequality:
\begin{align}
    &C_M(0)\simeq L^2(\log L+\gamma-\log2)/2\pi,\\
    &C_M(t)-C_M(0)\simeq(-i2\log2 )L^2t
\end{align}
for $M=L$ and $Mt/L\rightarrow 0$.
This superficial breakdown of the inequality comes from the fact that $A$ can depend on the Hamiltonian and the order operator \cite{Watanabe-Oshikawa}.
In the present case, 
the coefficients of the local terms in the order operator depend on the position $x$ that can be $\mathcal{O}(L)$, and $A$ becomes an increasing function of $L$.
The position fluctuation in a delocalized state is proportional to the system size, which leads to the macroscopic amplitude of the time-dependent correlation function.
Finally, we remark that one can investigate the same physics in the XX0 Heisenberg model.
Under the Jordan-Wigner transformation \cite{wigner1928paulische}, the Hamiltonian and the $z$-component spin are mapped to a one-dimensional free fermionic system and the particle density, respectively.


\begin{acknowledgements}
I thank Yuya O. Nakagawa for fruitful discussions.
This work was supported by JST CREST Grant No.~JPMJCR19T2, Japan. N.O. was supported by JSPS KAKENHI Grant No.~JP20K14373.
\end{acknowledgements}

\bibliography{Timecrystal}

\end{document}